\documentclass[11pt,a4paper]{article}
\usepackage[latin5]{inputenc}
\usepackage{amsmath}
\usepackage{amssymb}
\usepackage{graphicx}
\usepackage{subfigure}
\usepackage{caption}
\usepackage{float}
\usepackage{mathtools}
\usepackage{color}
\usepackage{cite}
\usepackage{mathrsfs}
\usepackage[]{geometry}
\usepackage{subfigure}

\newtheorem{theorem}{Theorem}
\title{Traveling Waves in Rational Expressions of Exponential Functions to the Conformable Time Fractional Jimbo-Miwa and Zakharov-Kuznetsov Equations}
\author{Alper Korkmaz$^{1,}$\thanks{korkmazalper@yandex.com}\, and Ozlem Ersoy Hepson$^{2}$\\%$^{,}$\thanks{Senior Author}\\
{\scriptsize $^{1}$ Çankırı Karatekin University, Department of Mathematics, 18200, Çankırı, Turkey.}\\
{\scriptsize $^{2}$ Eskişehir Osmangazi University, Department of Mathematics \& Computer, 26200, Eskişehir, Turkey.}
} 
\begin{document}
\maketitle
\begin{abstract}
The conformable time fractional Jimbo-Miwa and Zakharov-Kuznetsov equations are solved by the generalized form of the Kudryashov method. A simple compatible wave transformation is employed to reduce the dimension of the equations to one. The predicted solution is of the form of a rational expression of two finite series at both the numerator and the denominator. The terms of both series are of the powers of some functions having exponential expressions satisfying a particular ODE. The exact solutions are expressed explicitly in terms of powers of some exponential functions in form of rational expressions.  
\end{abstract}
%Traveling wave solutions-fractionalPDEs-KdV-likeeqns 
\section{Introduction}
The conformable time fractional Jimbo-Miwa (JM) equation of the form \cite{korkmaz2017a}
\begin{equation}
	u_{xxxy}+pu_yu_{xx}+qu_xu_{xy}+rT_t^{\beta}u_y-su_{xz}=0 \label{cfjm}
\end{equation}
,where $u$ is defined in $\mathbb{R}^3$, $t>0$ and the coefficients $p,q,r,s \in \mathbb{R}-\{0\}$, is considered. In this form of the equation, the conformal derivative operator $T_t^{\beta}$ represents the $\beta$.th order derivative ($\beta \in (0,1]$) with respect to the variable $t$. The integer ordered form of the JM equation (\ref{cfjm}) is a member of the KP-hierarchy and is not capable of passing the integrability tests\cite{jimbo1,cao1,xu1}. Details of the integrability conditions of the KP-hierarchy covering the JM equation was discussed in \cite{dorizzi1}.

Cao \cite{cao1} derived some non traveling exact solutions in forms of polynomials and logarithmic functions explicitly by implementing the methods of stable-range. Ma and Lee \cite{ma1} suggested some exact solutions covering some traveling wave forms, variable separated and lower degree polynomial functions of independent variables by using rational function transformation. Some solitary and periodic wave and variable separation solutions were determined by the improved mapping method\cite{ma2}. The generalized form of the $\tanh$-method is also capable of deriving some non-traveling-solitonic or traveling wave solution forms to the JM equation\cite{hong1}. Singh \cite{singh1} bilinearized the JM equation by a potential field to satisfy the connection to the binary Bell polynomial and defined a novel test function  that is a combination of some exponential,trigonometric and hyperbolic functions and to set up various solutions. The generalized wave solutions in rational exponential function forms with free parameters were described by the exp-function approach\cite{ozis1}. Some more exact solutions including solitons, exponential, trigonometric and functions of both wave solutions were obtained by applying the extended form of the homogenous balance approach\cite{liu1}. Kink, rational, periodic, and the solutions similar to solitons of the JM equation were constructed by the method of mapping based on Riccati equation\cite{li1}. 

The JM equation has also solutions in multi solution forms. Various types of multi-soliton solutions were constructed by the extended form of the homoclinic test method\cite{xu1}, the homogenous balance technique \cite{fang1}, Hirota's bilinear approach\cite{wazwaz1,hu1}, the general form of three-wave method\cite{li2}, Bäcklund transformation\cite{luu1}, ($G'/G$)-expansion and the separation of variables \cite{ming1,quan1}. Some fractional forms of the JM equation (in modified Riemann-Liouville and conformable derivative forms) have also been solved exactly by various methods such as modified and generalized Kudryashov, Riccati equation methods\cite{korkmaz2017a,kolebaje1,aksoy1}. 

The Zakharov-Kuznetsov (ZK) equation in conformable time fractional form in three space dimension given in \cite{korkmaz2017a} as
\begin{equation}
	T_{t}^{\beta}u+puu_x+qu_{zzz}+ru_{xxz}+su_{yyz}=0 \label{cfzk}
\end{equation}
is considered as the second equation to be solved exactly. The equation was derived to model stable $3$D ion-sound solitons with weak non linearity in a low pressure magnetized plasma\cite{zakharov1}. When the plasma contains $\kappa$-distributed hot and cold electrons in various temperatures, the ZK equation can also be obtained by using reductive perturbation techniques\cite{saini1}. Electron acoustic waves with weak non linearity were examined in a magnetic field for both magnetized and unmagnetized ions \cite{mace1}.

A general form ZK equation with arbitrarily chosen power non linearity was integrated by analysis of Lie symmetry,  the $\tanh$-function, ($G'/G$)-expansion and simple ansatz methods to construct cnoidal wave, singular and non-singular periodic, solitary wave and non topological soliton type exact solutions\cite{matebese1}. Some traveling wave type exact solutions were suggested by using exp-function and ($G'/G$)- and $F$-expansion approaches\cite{ebadi1}. Some more kink, antikink, solitary and periodic type traveling wave solutions were obtained the planar bifurcation theory\cite{zhang1}.

\section{Preliminaries and Essential Tools}
The $\beta$th order derivative in conformable sense is defined as
\begin{equation}
T_{t}^{\beta}(u(t))=\lim\limits_{\tau\rightarrow 0}{\dfrac{u(t+\tau t^{1-\beta})-u(t)}{\tau}},  \, \beta \in (0,1].
\end{equation}
in the positive half space for a function $u:[0,\infty)\rightarrow \mathbb{R}$\cite{khalil1}. The $\beta$th conformable derivative has the properties given below.

\begin{theorem}
Let $\beta \in (0,1]$, and assume that $u$ and $v$ are $\beta$-differentiable in the positive half plane (interval) $t>0$. Then,
\begin{itemize}
\item $T_t^{\beta}(au+bv)=aT_t^{\beta}(u)+bT_t^{\beta}(v)$ 
\item $T_t^{\beta}(t^p)=pt^{p-\beta}, \forall p \in \mathbb{R}$
\item $T_t^{\beta}(\lambda)=0$, for all constant function $u(t)=\lambda$
\item $T_t^{\beta}(uv)=uT_t^{\beta}(v)+vT_t^{\beta}(u)$
\item $T_t^{\beta}(\frac{u}{v})=\dfrac{vT_t^{\beta}(u)-uT_t^{\beta}(v)}{v^2}$
\item $T_t^{\beta}(u)(t)=t^{1-\beta}\frac{du}{dt}$
\end{itemize}
for $\forall a,b \in \mathbb{R}$\cite{atangana1,cenesiz1}.
\end{theorem}

The conformable derivative supports many significant properties like Laplace transform, exponential function, chain rule, Gronwall's inequality, various integration rules and Taylor series expansion\cite{abdeljawad1}. 
\begin{theorem}
Let $u$ be an $\beta$-differentiable function in conformable sense. Also suppose that $v$ is differentiable in classical sense and is defined in the range of $u$. Then,
\begin{equation}
T_t^{\beta}(u\circ v)(t)=t^{1-\beta}v^{\prime}(t)u^{\prime}(v(t))
\end{equation}
\end{theorem}

\section{Method of Solution}
A general non-linear PDE given as
\begin{equation}
F_1(u,T_t^{\beta}u,u_x,u_y,u_z,T_t^{2\beta}u,u_{xx},...)=0 \label{gfpde}
\end{equation}
where $u=u(x,y,z,t)$, $\beta \in (0,1]$ can be reduced to 
\begin{equation}
F_2(U,U{'},U{''},\ldots)=0 \label{gfode}
\end{equation}
where ($'$) indicates classical derivative of $U$ wrt $\omega$ by a simple transform of traveling wave
\begin{equation} 
u(x,y,z,t)=U(\omega), \omega = ax+by+cz-\frac{\nu}{\beta}t^{\beta} \label{wt}
\end{equation}
The compatible forms of the traveling wave transform were used in some recent studies\cite{korkmaz2017a,hosseini1,eslami1}. Implementing the classical balance procedure in the related terms gives the relation the numbers $M$ and $N$ required for the construction of the solution
\begin{equation}
	U(\omega)=\frac{\sum\limits_{i=0}^M{a_iP^i(\omega)}}{\sum\limits_{j=0}^N{b_jP^j(\omega)}}, \, a_M\neq 0, b_N \neq 0 \label{solt}
\end{equation}
where $P(\omega)$ satisfies the ODE
\begin{equation}
	\frac{dP}{d\omega}=\left(P^2-P\right)\ln{\left(A\right)} \label{simple}
\end{equation}
with the positive real $A\neq 1$. It should be noted that the solution of this ODE is
\begin{equation}
	P(\omega)=\frac{1}{1+\tilde{d}A^{\omega}}
\end{equation}
for a nonzero $\tilde{d}$. 
The solution procedure follows by substituting a more clear form of the solution (\ref{solt}) determined by using suitable values of $M$ and $N$ into (\ref{gfode}). The coefficients of powers of $P$ in the resultant polynomial is forced to be zero. Thus, an algebraic system of equations is constructed. The solution of this system gives the relation between the parameters used in the target equation and the wave transform. Once the relations between the parameters are determined the solution to (\ref{gfode}) can be expressed explicitly. The final step of the procedure is to express the solutions of the target fractional PDE in terms of the original variables.

\section{The solutions of the conformable time fractional JM Equation}
The wave transform (\ref{wt}) reduces the JM equation (after integrating the resultant ODE once) to
\begin{equation}
	{a}^{3}b{\frac {{\rm d}^{3}}{{\rm d}{\omega}^{3}}}U\left( \omega \right) +\frac{1}{2}\,{a}^{2}b \left( p+q \right)  \left( {\frac {\rm d}{{\rm d}\omega}}U
 \left( \omega \right)  \right) ^{2}+ \left( -sac+\nu\,br \right) {\frac 
{\rm d}{{\rm d}\omega}}U \left( \omega \right) =K \label{odejm}
\end{equation}
 where $K$ is the constant of integration. Balancing $U'''$ and $(U')^2$ gives the relation $M=N+1$. Choose $M=2$ and $N=1$. Then, the solution (\ref{solt}) is expressed as
\begin{equation}
	U(\omega)=\frac{\sum\limits_{i=0}^2{a_iP^i(\omega)}}{\sum\limits_{j=0}^1{b_jP^j(\omega)}}=\frac{a_0+a_1P+a_2P^2}{b_0+b_1P}
\end{equation}
with $a_2\neq 0$ and $b_1\neq 0$. Substituting this solution into (\ref{odejm}) gives
{\scriptsize
\begin{equation}
\begin{aligned}
	&{a}^{3}b \left( {\frac {a_{{1}}{\frac {{\rm d}^{3}}{{\rm d}{\omega}^{3}}}
P   +6\,a_{{2}} \left( {\frac {\rm d}{{\rm d}\omega}}P
    \right) {\frac {{\rm d}^{2}}{{\rm d}{\omega}^{2}}}P
   +2\,a_{{2}}P   {\frac {{\rm d}^{3
}}{{\rm d}{\omega}^{3}}}P   }{b_{{0}}+b_{{1}}P }}\right. \\
&\left. -3\,{\frac { \left( a_{{1}}{\frac {{\rm d}^{2}}{{\rm d}{
\omega}^{2}}}P   +2\,a_{{2}} \left( {\frac {\rm d}{
{\rm d}\omega}}P    \right) ^{2}+2\,a_{{2}}P{\frac {{\rm d}^{2}}{{\rm d}{\omega}^{2}}}P   
 \right) b_{{1}}{\frac {\rm d}{{\rm d}\omega}}P   }{
 \left( b_{{0}}+b_{{1}}P    \right) ^{2}}}+6\,{\frac 
{ \left( a_{{1}}{\frac {\rm d}{{\rm d}\omega}}P   +2\,a_
{{2}}P   {\frac {\rm d}{{\rm d}\omega}}P \right) {b_{{1}}}^{2} \left( {\frac {\rm d}{{\rm d}\omega}}P
    \right) ^{2}}{ \left( b_{{0}}+b_{{1}}P   \right) ^{3}}}\right. \\
&\left. -3\,{\frac { \left( a_{{1}}{\frac {\rm d}{
{\rm d}\omega}}P   +2\,a_{{2}}P   {
\frac {\rm d}{{\rm d}\omega}}P    \right) b_{{1}}{\frac 
{{\rm d}^{2}}{{\rm d}{\omega}^{2}}}P   }{ \left( b_{{0}}
+b_{{1}}P    \right) ^{2}}}-6\,{\frac { \left( a_{{0}
}+a_{{1}}P   +a_{{2}} P^{2} \right) {b_{{1}}}^{3} \left( {\frac {\rm d}{{\rm d}\omega}}
P    \right) ^{3}}{ \left( b_{{0}}+b_{{1}}P   \right) ^{4}}}\right. \\
&\left. +6\,{\frac { \left( a_{{0}}+a_{{1}}P
   +a_{{2}}P ^{2}
 \right) {b_{{1}}}^{2} \left( {\frac {\rm d}{{\rm d}\omega}}P  \right) {\frac {{\rm d}^{2}}{{\rm d}{\omega}^{2}}}P  }{ \left( b_{{0}}+b_{{1}}P    \right) ^{3}}}
-{\frac { \left( a_{{0}}+a_{{1}}P   +a_{{2}} P^{2} \right) b_{{1}}{\frac {{\rm d}^{3}}{
{\rm d}{\omega}^{3}}}P   }{ \left( b_{{0}}+b_{{1}}P
    \right) ^{2}}} \right) \\
&+1/2\,{a}^{2}b \left( p+q
 \right)  \left( {\frac {a_{{1}}{\frac {\rm d}{{\rm d}\omega}}P +2\,a_{{2}}P   {\frac {\rm d}{{\rm d}\omega}
}P   }{b_{{0}}+b_{{1}}P   }}-{\frac 
{ \left( a_{{0}}+a_{{1}}P   +a_{{2}} \left( P \left( 
\omega \right)  \right) ^{2} \right) b_{{1}}{\frac {\rm d}{{\rm d}\omega}}P
   }{ \left( b_{{0}}+b_{{1}}P   
 \right) ^{2}}} \right) ^{2}\\
&+ \left( -acs+b\nu\,r \right)  \left( {
\frac {a_{{1}}{\frac {\rm d}{{\rm d}\omega}}P   +2\,a_{{
2}}P   {\frac {\rm d}{{\rm d}\omega}}P }{b_{{0}}+b_{{1}}P   }}-{\frac { \left( a_{{0
}}+a_{{1}}P   +a_{{2}} \left( P   
 \right) ^{2} \right) b_{{1}}{\frac {\rm d}{{\rm d}\omega}}P }{ \left( b_{{0}}+b_{{1}}P    \right) ^{2}}}
 \right) =K
\end{aligned}
\end{equation}}
This equation can be rearranged by using (\ref{simple}). Then, the coefficients of powers of $P$ is equalized to zero to give an algebraic system of equations
{\scriptsize
\begin{equation}
\begin{aligned}
-K{b_{{0}}}^{4}&=0\\
{a}^{3}ba_{{0}}{b_{{0}}}^{2}b_{{1}} \left( \ln  \left( A \right) 
 \right) ^{3}-{a}^{3}ba_{{1}}{b_{{0}}}^{3} \left( \ln  \left( A
 \right)  \right) ^{3}-saca_{{0}}{b_{{0}}}^{2}b_{{1}}\ln  \left( A
 \right) \\
+saca_{{1}}{b_{{0}}}^{3}\ln  \left( A \right) +\nu\,bra_{{0}}
{b_{{0}}}^{2}b_{{1}}\ln  \left( A \right) -\nu\,bra_{{1}}{b_{{0}}}^{3}
\ln  \left( A \right) -4\,K{b_{{0}}}^{3}b_{{1}}&=0\\
%%%%%%%%%%%%%%%%%%%%%%%%%%%%%%%%%%%%%
-7\,{a}^{3}ba_{{0}}{b_{{0}}}^{2}b_{{1}} \left( \ln  \left( A \right) 
 \right) ^{3}-4\,{a}^{3}ba_{{0}}b_{{0}}{b_{{1}}}^{2} \left( \ln 
 \left( A \right)  \right) ^{3}+7\,{a}^{3}ba_{{1}}{b_{{0}}}^{3}
 \left( \ln  \left( A \right)  \right) ^{3}\\
+4\,{a}^{3}ba_{{1}}{b_{{0}}
}^{2}b_{{1}} \left( \ln  \left( A \right)  \right) ^{3}-8\,{a}^{3}ba_{
{2}}{b_{{0}}}^{3} \left( \ln  \left( A \right)  \right) ^{3}+1/2\,{a}^
{2}bp{a_{{0}}}^{2}{b_{{1}}}^{2} \left( \ln  \left( A \right)  \right) 
^{2}-{a}^{2}bpa_{{0}}a_{{1}}b_{{0}}b_{{1}} \left( \ln  \left( A
 \right)  \right) ^{2}\\
+1/2\,{a}^{2}bp{a_{{1}}}^{2}{b_{{0}}}^{2}
 \left( \ln  \left( A \right)  \right) ^{2}+1/2\,{a}^{2}bq{a_{{0}}}^{2
}{b_{{1}}}^{2} \left( \ln  \left( A \right)  \right) ^{2}-{a}^{2}bqa_{
{0}}a_{{1}}b_{{0}}b_{{1}} \left( \ln  \left( A \right)  \right) ^{2}+1
/2\,{a}^{2}bq{a_{{1}}}^{2}{b_{{0}}}^{2} \left( \ln  \left( A \right) 
 \right) ^{2}\\
+saca_{{0}}{b_{{0}}}^{2}b_{{1}}\ln  \left( A \right) -2\,
saca_{{0}}b_{{0}}{b_{{1}}}^{2}\ln  \left( A \right) -saca_{{1}}{b_{{0}
}}^{3}\ln  \left( A \right) +2\,saca_{{1}}{b_{{0}}}^{2}b_{{1}}\ln 
 \left( A \right) +2\,saca_{{2}}{b_{{0}}}^{3}\ln  \left( A \right) \\-
\nu\,bra_{{0}}{b_{{0}}}^{2}b_{{1}}\ln  \left( A \right) +2\,\nu\,bra_{
{0}}b_{{0}}{b_{{1}}}^{2}\ln  \left( A \right) +\nu\,bra_{{1}}{b_{{0}}}
^{3}\ln  \left( A \right) -2\,\nu\,bra_{{1}}{b_{{0}}}^{2}b_{{1}}\ln 
 \left( A \right) \\
-2\,\nu\,bra_{{2}}{b_{{0}}}^{3}\ln  \left( A
 \right) -6\,K{b_{{0}}}^{2}{b_{{1}}}^{2}&=0\\
%%%%%%%%%%%%%%%%%%%%
saca_{{1}}b_{{0}}{b_{{1}}}^{2}\ln  \left( A \right) +5\,saca_{{2}}{b_{
{0}}}^{2}b_{{1}}\ln  \left( A \right) -\nu\,bra_{{1}}b_{{0}}{b_{{1}}}^
{2}\ln  \left( A \right) -5\,\nu\,bra_{{2}}{b_{{0}}}^{2}b_{{1}}\ln 
 \left( A \right) \\
+2\,{a}^{2}bqa_{{0}}a_{{1}}b_{{0}}b_{{1}} \left( 
\ln  \left( A \right)  \right) ^{2}+2\,{a}^{2}bpa_{{0}}a_{{1}}b_{{0}}b
_{{1}} \left( \ln  \left( A \right)  \right) ^{2}-4\,Kb_{{0}}{b_{{1}}}
^{3}-2\,saca_{{2}}{b_{{0}}}^{3}\ln  \left( A \right) \\
+2\,\nu\,bra_{{2}
}{b_{{0}}}^{3}\ln  \left( A \right) -{a}^{2}bp{a_{{0}}}^{2}{b_{{1}}}^{
2} \left( \ln  \left( A \right)  \right) ^{2}-{a}^{2}bp{a_{{1}}}^{2}{b
_{{0}}}^{2} \left( \ln  \left( A \right)  \right) ^{2}\\
-{a}^{2}bq{a_{{0
}}}^{2}{b_{{1}}}^{2} \left( \ln  \left( A \right)  \right) ^{2}-{a}^{2
}bq{a_{{1}}}^{2}{b_{{0}}}^{2} \left( \ln  \left( A \right)  \right) ^{
2}+12\,{a}^{3}ba_{{0}}{b_{{0}}}^{2}b_{{1}} \left( \ln  \left( A
 \right)  \right) ^{3}+10\,{a}^{3}ba_{{0}}b_{{0}}{b_{{1}}}^{2} \left( 
\ln  \left( A \right)  \right) ^{3}\\
-10\,{a}^{3}ba_{{1}}{b_{{0}}}^{2}b_
{{1}} \left( \ln  \left( A \right)  \right) ^{3}+{a}^{3}ba_{{0}}{b_{{1
}}}^{3} \left( \ln  \left( A \right)  \right) ^{3}+38\,{a}^{3}ba_{{2}}
{b_{{0}}}^{3} \left( \ln  \left( A \right)  \right) ^{3}\\
-12\,{a}^{3}ba
_{{1}}{b_{{0}}}^{3} \left( \ln  \left( A \right)  \right) ^{3}-2\,\nu
\,bra_{{0}}b_{{0}}{b_{{1}}}^{2}\ln  \left( A \right) +2\,\nu\,bra_{{1}
}{b_{{0}}}^{2}b_{{1}}\ln  \left( A \right) +2\,saca_{{0}}b_{{0}}{b_{{1
}}}^{2}\ln  \left( A \right) -2\,saca_{{1}}{b_{{0}}}^{2}b_{{1}}\ln 
 \left( A \right) \\
+2\,{a}^{2}bpa_{{1}}a_{{2}}{b_{{0}}}^{2} \left( \ln 
 \left( A \right)  \right) ^{2}+2\,{a}^{2}bqa_{{1}}a_{{2}}{b_{{0}}}^{2
} \left( \ln  \left( A \right)  \right) ^{2}-saca_{{0}}{b_{{1}}}^{3}
\ln  \left( A \right) +\nu\,bra_{{0}}{b_{{1}}}^{3}\ln  \left( A
 \right) -{a}^{3}ba_{{1}}b_{{0}}{b_{{1}}}^{2} \left( \ln  \left( A
 \right)  \right) ^{3}\\
-5\,{a}^{3}ba_{{2}}{b_{{0}}}^{2}b_{{1}} \left( 
\ln  \left( A \right)  \right) ^{3}-2\,{a}^{2}bqa_{{0}}a_{{2}}b_{{0}}b
_{{1}} \left( \ln  \left( A \right)  \right) ^{2}-2\,{a}^{2}bpa_{{0}}a
_{{2}}b_{{0}}b_{{1}} \left( \ln  \left( A \right)  \right) ^{2}&=0\\
%%%%%%%%%%%%%%%%%%%%%%%%%%
4\,saca_{{2}}b_{{0}}{b_{{1}}}^{2}\ln  \left( A \right) -4\,\nu\,bra_{{
2}}b_{{0}}{b_{{1}}}^{2}\ln  \left( A \right) -saca_{{1}}b_{{0}}{b_{{1}
}}^{2}\ln  \left( A \right) -5\,saca_{{2}}{b_{{0}}}^{2}b_{{1}}\ln 
 \left( A \right) \\
+\nu\,bra_{{1}}b_{{0}}{b_{{1}}}^{2}\ln  \left( A
 \right) +5\,\nu\,bra_{{2}}{b_{{0}}}^{2}b_{{1}}\ln  \left( A \right) -
{a}^{2}bqa_{{0}}a_{{1}}b_{{0}}b_{{1}} \left( \ln  \left( A \right) 
 \right) ^{2}-{a}^{2}bpa_{{0}}a_{{1}}b_{{0}}b_{{1}} \left( \ln 
 \left( A \right)  \right) ^{2}\\
-K{b_{{1}}}^{4}+1/2\,{a}^{2}bp{a_{{0}}}
^{2}{b_{{1}}}^{2} \left( \ln  \left( A \right)  \right) ^{2}+1/2\,{a}^
{2}bp{a_{{1}}}^{2}{b_{{0}}}^{2} \left( \ln  \left( A \right)  \right) 
^{2}+1/2\,{a}^{2}bq{a_{{0}}}^{2}{b_{{1}}}^{2} \left( \ln  \left( A
 \right)  \right) ^{2}\\
+1/2\,{a}^{2}bq{a_{{1}}}^{2}{b_{{0}}}^{2}
 \left( \ln  \left( A \right)  \right) ^{2}-6\,{a}^{3}ba_{{0}}{b_{{0}}
}^{2}b_{{1}} \left( \ln  \left( A \right)  \right) ^{3}-6\,{a}^{3}ba_{
{0}}b_{{0}}{b_{{1}}}^{2} \left( \ln  \left( A \right)  \right) ^{3}+6
\,{a}^{3}ba_{{1}}{b_{{0}}}^{2}b_{{1}} \left( \ln  \left( A \right) 
 \right) ^{3}\\
-{a}^{3}ba_{{0}}{b_{{1}}}^{3} \left( \ln  \left( A
 \right)  \right) ^{3}-54\,{a}^{3}ba_{{2}}{b_{{0}}}^{3} \left( \ln 
 \left( A \right)  \right) ^{3}+6\,{a}^{3}ba_{{1}}{b_{{0}}}^{3}
 \left( \ln  \left( A \right)  \right) ^{3}\\
-{a}^{2}bpa_{{0}}a_{{2}}{b_
{{1}}}^{2} \left( \ln  \left( A \right)  \right) ^{2}-{a}^{2}bqa_{{0}}
a_{{2}}{b_{{1}}}^{2} \left( \ln  \left( A \right)  \right) ^{2}-4\,{a}
^{2}bpa_{{1}}a_{{2}}{b_{{0}}}^{2} \left( \ln  \left( A \right) 
 \right) ^{2}-4\,{a}^{2}bqa_{{1}}a_{{2}}{b_{{0}}}^{2} \left( \ln 
 \left( A \right)  \right) ^{2}\\
+saca_{{0}}{b_{{1}}}^{3}\ln  \left( A
 \right) -\nu\,bra_{{0}}{b_{{1}}}^{3}\ln  \left( A \right) +{a}^{3}ba_
{{1}}b_{{0}}{b_{{1}}}^{2} \left( \ln  \left( A \right)  \right) ^{3}+
41\,{a}^{3}ba_{{2}}{b_{{0}}}^{2}b_{{1}} \left( \ln  \left( A \right) 
 \right) ^{3}\\
-4\,{a}^{3}ba_{{2}}b_{{0}}{b_{{1}}}^{2} \left( \ln 
 \left( A \right)  \right) ^{3}+2\,{a}^{2}bp{a_{{2}}}^{2}{b_{{0}}}^{2}
 \left( \ln  \left( A \right)  \right) ^{2}+2\,{a}^{2}bq{a_{{2}}}^{2}{
b_{{0}}}^{2} \left( \ln  \left( A \right)  \right) ^{2}\\
+{a}^{2}bqa_{{1
}}a_{{2}}b_{{0}}b_{{1}} \left( \ln  \left( A \right)  \right) ^{2}+{a}
^{2}bpa_{{1}}a_{{2}}b_{{0}}b_{{1}} \left( \ln  \left( A \right) 
 \right) ^{2}+4\,{a}^{2}bqa_{{0}}a_{{2}}b_{{0}}b_{{1}} \left( \ln 
 \left( A \right)  \right) ^{2}+4\,{a}^{2}bpa_{{0}}a_{{2}}b_{{0}}b_{{1
}} \left( \ln  \left( A \right)  \right) ^{2}&=0\\
%%%%%%%%%%%%%%%%%%%%%%%%%%%%%
24\,{a}^{3}ba_{{2}}{b_{{0}}}^{3} \left( \ln  \left( A \right) 
 \right) ^{3}-72\,{a}^{3}ba_{{2}}{b_{{0}}}^{2}b_{{1}} \left( \ln 
 \left( A \right)  \right) ^{3}+28\,{a}^{3}ba_{{2}}b_{{0}}{b_{{1}}}^{2
} \left( \ln  \left( A \right)  \right) ^{3}\\
-{a}^{3}ba_{{2}}{b_{{1}}}^
{3} \left( \ln  \left( A \right)  \right) ^{3}-2\,{a}^{2}bpa_{{0}}a_{{
2}}b_{{0}}b_{{1}} \left( \ln  \left( A \right)  \right) ^{2}+2\,{a}^{2
}bpa_{{0}}a_{{2}}{b_{{1}}}^{2} \left( \ln  \left( A \right)  \right) ^
{2}+2\,{a}^{2}bpa_{{1}}a_{{2}}{b_{{0}}}^{2} \left( \ln  \left( A
 \right)  \right) ^{2}\\
-2\,{a}^{2}bpa_{{1}}a_{{2}}b_{{0}}b_{{1}}
 \left( \ln  \left( A \right)  \right) ^{2}-4\,{a}^{2}bp{a_{{2}}}^{2}{
b_{{0}}}^{2} \left( \ln  \left( A \right)  \right) ^{2}+2\,{a}^{2}bp{a
_{{2}}}^{2}b_{{0}}b_{{1}} \left( \ln  \left( A \right)  \right) ^{2}-2
\,{a}^{2}bqa_{{0}}a_{{2}}b_{{0}}b_{{1}} \left( \ln  \left( A \right) 
 \right) ^{2}\\
+2\,{a}^{2}bqa_{{0}}a_{{2}}{b_{{1}}}^{2} \left( \ln 
 \left( A \right)  \right) ^{2}+2\,{a}^{2}bqa_{{1}}a_{{2}}{b_{{0}}}^{2
} \left( \ln  \left( A \right)  \right) ^{2}-2\,{a}^{2}bqa_{{1}}a_{{2}
}b_{{0}}b_{{1}} \left( \ln  \left( A \right)  \right) ^{2}-4\,{a}^{2}b
q{a_{{2}}}^{2}{b_{{0}}}^{2} \left( \ln  \left( A \right)  \right) ^{2}
\\
+2\,{a}^{2}bq{a_{{2}}}^{2}b_{{0}}b_{{1}} \left( \ln  \left( A \right) 
 \right) ^{2}-4\,saca_{{2}}b_{{0}}{b_{{1}}}^{2}\ln  \left( A \right) +
saca_{{2}}{b_{{1}}}^{3}\ln  \left( A \right) +4\,\nu\,bra_{{2}}b_{{0}}
{b_{{1}}}^{2}\ln  \left( A \right) -\nu\,bra_{{2}}{b_{{1}}}^{3}\ln 
 \left( A \right)&=0 \\
%%%%%%%%%%%%%%%%%%%%%%%%%%%%%%%%%%%
36\,{a}^{3}ba_{{2}}{b_{{0}}}^{2}b_{{1}} \left( \ln  \left( A \right) 
 \right) ^{3}-48\,{a}^{3}ba_{{2}}b_{{0}}{b_{{1}}}^{2} \left( \ln 
 \left( A \right)  \right) ^{3}+7\,{a}^{3}ba_{{2}}{b_{{1}}}^{3}
 \left( \ln  \left( A \right)  \right) ^{3}-{a}^{2}bpa_{{0}}a_{{2}}{b_
{{1}}}^{2} \left( \ln  \left( A \right)  \right) ^{2}\\
+{a}^{2}bpa_{{1}}
a_{{2}}b_{{0}}b_{{1}} \left( \ln  \left( A \right)  \right) ^{2}+2\,{a
}^{2}bp{a_{{2}}}^{2}{b_{{0}}}^{2} \left( \ln  \left( A \right) 
 \right) ^{2}-4\,{a}^{2}bp{a_{{2}}}^{2}b_{{0}}b_{{1}} \left( \ln 
 \left( A \right)  \right) ^{2}+1/2\,{a}^{2}bp{a_{{2}}}^{2}{b_{{1}}}^{
2} \left( \ln  \left( A \right)  \right) ^{2}\\
-{a}^{2}bqa_{{0}}a_{{2}}{
b_{{1}}}^{2} \left( \ln  \left( A \right)  \right) ^{2}+{a}^{2}bqa_{{1
}}a_{{2}}b_{{0}}b_{{1}} \left( \ln  \left( A \right)  \right) ^{2}+2\,
{a}^{2}bq{a_{{2}}}^{2}{b_{{0}}}^{2} \left( \ln  \left( A \right) 
 \right) ^{2}-4\,{a}^{2}bq{a_{{2}}}^{2}b_{{0}}b_{{1}} \left( \ln 
 \left( A \right)  \right) ^{2}\\
+1/2\,{a}^{2}bq{a_{{2}}}^{2}{b_{{1}}}^{
2} \left( \ln  \left( A \right)  \right) ^{2}-saca_{{2}}{b_{{1}}}^{3}
\ln  \left( A \right) +\nu\,bra_{{2}}{b_{{1}}}^{3}\ln  \left( A
 \right)&=0\\
%%%%%%%%%%%%%%%%%%%%%%%%%%%%%%%%%%%
24\,{a}^{3}ba_{{2}}b_{{0}}{b_{{1}}}^{2} \left( \ln  \left( A \right) 
 \right) ^{3}-12\,{a}^{3}ba_{{2}}{b_{{1}}}^{3} \left( \ln  \left( A
 \right)  \right) ^{3}+2\,{a}^{2}bp{a_{{2}}}^{2}b_{{0}}b_{{1}} \left( 
\ln  \left( A \right)  \right) ^{2}\\
-{a}^{2}bp{a_{{2}}}^{2}{b_{{1}}}^{2
} \left( \ln  \left( A \right)  \right) ^{2}+2\,{a}^{2}bq{a_{{2}}}^{2}
b_{{0}}b_{{1}} \left( \ln  \left( A \right)  \right) ^{2}-{a}^{2}bq{a_
{{2}}}^{2}{b_{{1}}}^{2} \left( \ln  \left( A \right)  \right) ^{2}&=0\\
%%%%%%%%%%%%%%%%%%%%%%%%%%%%%%%%%%
6\,{a}^{3}ba_{{2}}{b_{{1}}}^{3} \left( \ln  \left( A \right)  \right) 
^{3}+1/2\,{a}^{2}bp{a_{{2}}}^{2}{b_{{1}}}^{2} \left( \ln  \left( A
 \right)  \right) ^{2}+1/2\,{a}^{2}bq{a_{{2}}}^{2}{b_{{1}}}^{2}
 \left( \ln  \left( A \right)  \right) ^{2}&=0
%%%%%%%%%%%%%%%%%%%%%%%%%%%%%%%%%
 \label{sysjm}
\end{aligned}
\end{equation}}
The solution of this system (\ref{sysjm}) for $a_0$, $a_1$, $a_2$, $b_0$, $b_1$, $\nu$ and $K$ gives
\begin{equation}
	\begin{aligned}
	 a_0&= -\frac{a_1}{2} \\
	 %a_1&= \\
	 a_2&=-12\,{\frac {\ln  \left( A \right) ab_{{1}}}{p+q}} \\
	 b_0&= -\frac{b_1}{2}\\
	 %b_1&= \\
	\nu&= -{\frac {a \left( 4\, \left( \ln  \left( A \right)  \right) ^{2}{a}^{2}
b-cs \right) }{br}}
\\
	K&=0
	\end{aligned}
\end{equation}
for arbitrarily chosen $a$, $b$, $c$, $a_1$ and $b_1$. Thus, the solution to (\ref{odejm}) is determined as
\begin{equation}
	U_1(\omega)={ \left( -1/2\,a_{{1}}+{\frac {a_{{1}}}{1+\tilde{d}{A}^{\omega}}}-12\,{\frac {
\ln  \left( A \right) ab_{{1}}}{ \left( p+q \right)  \left( 1+d{A}^{
\omega} \right) ^{2}}} \right)  \left( -1/2\,b_{{1}}+{\frac {b_{{1}}}{1+\tilde{d}
{A}^{\omega}}} \right) ^{-1}}
\end{equation}
Returning the original variables $(x,y,z,t)$ gives the solution to the conformable time fractional JM equation (\ref{cfjm}) as
\begin{equation}
\begin{aligned}
	u_1(x,y,z,t)=\frac{ \left( -\frac{1}{2}\,a_{{1}}+{\frac {a_{{1}}}{1+\tilde{d}{A}^{ax+by+cz+{\frac {a \left( 4\, \left( \ln  \left( A \right)  \right) ^{2}{a}^{2}
b-cs \right) }{br}}\frac{t^{\beta}}{\beta}}}}-12\,{\frac {
\ln  \left( A \right) ab_{{1}}}{ \left( p+q \right)  \left( 1+\tilde{d}{A}^{
{ax+by+cz+{\frac {a \left( 4\, \left( \ln  \left( A \right)  \right) ^{2}{a}^{2}
b-cs \right) }{br}}\frac{t^{\beta}}{\beta}}} \right) ^{2}}} \right)}{\left( -1/2\,b_{{1}}+{\frac {b_{{1}}}{1+\tilde{d}
{A}^{ax+by+cz+{\frac {a \left( 4\, \left( \ln  \left( A \right)  \right) ^{2}{a}^{2}
b-cs \right) }{br}}\frac{t^{\beta}}{\beta}}}} \right)}
\end{aligned}
\end{equation}
The algebraic system (\ref{sysjm}) has one more solution as
\begin{equation}
	\begin{aligned}
	 a_0&= {\frac {b_{{0}} \left( 12\,\ln  \left( A \right) ab_{{0}}+pa_{{1}}+qa_
{{1}} \right) }{b_{{1}} \left( p+q \right) }} \\
	 %a_1&= \\
	 a_2&=-12\,{\frac {\ln  \left( A \right) ab_{{1}}}{p+q}} \\
	 %b_0&= -\frac{b_1}{2}\\
	 %b_1&= \\
	\nu&= -{\frac {a \left(  \left( \ln  \left( A \right)  \right) ^{2}{a}^{2}b-c
s \right) }{br}}\\
	K&=0
	\end{aligned}
\end{equation}
for arbitrary $a_1$, $b_0$, $b_1$, $a$, $b$ and $c$. Then, the solution to (\ref{odejm}) is expressed as
\begin{equation}
	U_2(\omega)= \frac{ \left( {\frac {b_{{0}} \left( 12\,\ln  \left( A \right) ab_{{0}}+pa
_{{1}}+qa_{{1}} \right) }{b_{{1}} \left( p+q \right) }}+{\frac {a_{{1}
}}{1+\tilde{d}{A}^{\omega}}}-12\,{\frac {\ln  \left( A \right) ab_{{1}}}{ \left( 
p+q \right)  \left( 1+\tilde{d}{A}^{\omega} \right) ^{2}}} \right) }{ \left( b_{{0}
}+{\frac {b_{{1}}}{1+\tilde{d}{A}^{\omega}}} \right) }
\end{equation}
Thus, the solution to the conformable time fractional JM equation (\ref{cfjm}) is constructed as
{\scriptsize
\begin{equation}
	u_2(x,y,z,t)= \frac{ \left( {\frac {b_{{0}} \left( 12\,\ln  \left( A \right) ab_{{0}}+pa
_{{1}}+qa_{{1}} \right) }{b_{{1}} \left( p+q \right) }}+{\frac {a_{{1}
}}{1+\tilde{d}{A}^{ax+by+cz+{\frac {a \left(  \left( \ln  \left( A \right)  \right) ^{2}{a}^{2}b-c
s \right) }{br}}\frac{t^{\beta}}{\beta}}}}-12\,{\frac {\ln  \left( A \right) ab_{{1}}}{ \left( 
p+q \right)  \left( 1+\tilde{d}{A}^{ax+by+cz+{\frac {a \left(  \left( \ln  \left( A \right)  \right) ^{2}{a}^{2}b-c
s \right) }{br}}\frac{t^{\beta}}{\beta}} \right) ^{2}}} \right) }{ \left( b_{{0}
}+{\frac {b_{{1}}}{1+\tilde{d}{A}^{ax+by+cz+{\frac {a \left(  \left( \ln  \left( A \right)  \right) ^{2}{a}^{2}b-c
s \right) }{br}}\frac{t^{\beta}}{\beta}}}} \right) }
\end{equation}}
\section{The solutions of the conformable time fractional ZK Equation}
The traveling wave transform (\ref{wt}) reduces the conformable time fractional ZK equation (\ref{cfzk}) to
\begin{equation}
	-\nu\,U \left( \omega \right) +\frac{1}{2}\,ap \left( U \left( \omega \right) 
 \right) ^{2}+ \left( {a}^{2}br+{b}^{2}cs+{c}^{3}q \right) {\frac {
{\rm d}^{2}}{{\rm d}{\omega}^{2}}}U \left( \omega \right) =K,\, K \, constant \label{zkode}
\end{equation}
after integrating the resultant ODE once. Balancing $U^2$ and $U''$ gives the relation $M=N+2$. Assuming $N=1$ and $M=3$ gives the predicted solution of the form
\begin{equation}
	U(\omega)=\frac{a_0+a_1P+a_2P^2+a_3P^3}{b_0+b_1P}
\end{equation}
where $a_3\neq 0$ and $b_1 \neq 0$. Substituting this solution into (\ref{zkode}) and arranging the resultant equation gives

{\scriptsize
\begin{equation}
	\begin{aligned}
&\left( 6\, \left( \ln  \left( A \right)  \right) ^{2}{a}^{2}bra_{{3}}
{b_{{1}}}^{2}+6\, \left( \ln  \left( A \right)  \right) ^{2}{b}^{2}csa
_{{3}}{b_{{1}}}^{2}+1/2\,ap{a_{{3}}}^{2}b_{{1}}+6\, \left( \ln 
 \left( A \right)  \right) ^{2}{c}^{3}qa_{{3}}{b_{{1}}}^{2} \right) 
 P ^{7}\left( \omega \right) \\
&+ \left( 2\, \left( \ln 
 \left( A \right)  \right) ^{2}{a}^{2}bra_{{2}}{b_{{1}}}^{2}-10\,
 \left( \ln  \left( A \right)  \right) ^{2}{a}^{2}bra_{{3}}{b_{{1}}}^{
2}+2\, \left( \ln  \left( A \right)  \right) ^{2}{b}^{2}csa_{{2}}{b_{{
1}}}^{2}-10\, \left( \ln  \left( A \right)  \right) ^{2}{b}^{2}csa_{{3
}}{b_{{1}}}^{2}+16\, \left( \ln  \left( A \right)  \right) ^{2}{c}^{3}
qa_{{3}}b_{{0}}b_{{1}}\right. \\
&\left. +1/2\,ap{a_{{3}}}^{2}b_{{0}}+apa_{{2}}a_{{3}}b_{
{1}}+2\, \left( \ln  \left( A \right)  \right) ^{2}{c}^{3}qa_{{2}}{b_{
{1}}}^{2}-10\, \left( \ln  \left( A \right)  \right) ^{2}{c}^{3}qa_{{3
}}{b_{{1}}}^{2}+16\, \left( \ln  \left( A \right)  \right) ^{2}{a}^{2}
bra_{{3}}b_{{0}}b_{{1}}+16\, \left( \ln  \left( A \right)  \right) ^{2
}{b}^{2}csa_{{3}}b_{{0}}b_{{1}} \right) P ^{6}\left( \omega  \right) \\
&+ \left( -\nu\,a_{{3}}{b_{{1}}}^{2}-3\, \left( \ln 
 \left( A \right)  \right) ^{2}{a}^{2}bra_{{2}}{b_{{1}}}^{2}+12\,
 \left( \ln  \left( A \right)  \right) ^{2}{a}^{2}bra_{{3}}{b_{{0}}}^{
2}+4\, \left( \ln  \left( A \right)  \right) ^{2}{a}^{2}bra_{{3}}{b_{{
1}}}^{2}-3\, \left( \ln  \left( A \right)  \right) ^{2}{b}^{2}csa_{{2}
}{b_{{1}}}^{2}\right. \\
&\left. +12\, \left( \ln  \left( A \right)  \right) ^{2}{b}^{2}c
sa_{{3}}{b_{{0}}}^{2}+4\, \left( \ln  \left( A \right)  \right) ^{2}{b
}^{2}csa_{{3}}{b_{{1}}}^{2}+6\, \left( \ln  \left( A \right)  \right) 
^{2}{c}^{3}qa_{{2}}b_{{0}}b_{{1}}-27\, \left( \ln  \left( A \right) 
 \right) ^{2}{c}^{3}qa_{{3}}b_{{0}}b_{{1}}+1/2\,ap{a_{{2}}}^{2}b_{{1}}
+apa_{{1}}a_{{3}}b_{{1}}\right. \\
&\left. +apa_{{2}}a_{{3}}b_{{0}}-3\, \left( \ln 
 \left( A \right)  \right) ^{2}{c}^{3}qa_{{2}}{b_{{1}}}^{2}+12\,
 \left( \ln  \left( A \right)  \right) ^{2}{c}^{3}qa_{{3}}{b_{{0}}}^{2
}\right. \\
&\left. +4\, \left( \ln  \left( A \right)  \right) ^{2}{c}^{3}qa_{{3}}{b_{{1}
}}^{2}+6\, \left( \ln  \left( A \right)  \right) ^{2}{a}^{2}bra_{{2}}b
_{{0}}b_{{1}}-27\, \left( \ln  \left( A \right)  \right) ^{2}{a}^{2}br
a_{{3}}b_{{0}}b_{{1}}+6\, \left( \ln  \left( A \right)  \right) ^{2}{b
}^{2}csa_{{2}}b_{{0}}b_{{1}}-27\, \left( \ln  \left( A \right) 
 \right) ^{2}{b}^{2}csa_{{3}}b_{{0}}b_{{1}} \right)   P ^{5}\left( 
\omega \right)\\
&+ \left( -\nu\,a_{{2}}{b_{{1}}}^{2}+6\,
 \left( \ln  \left( A \right)  \right) ^{2}{b}^{2}csa_{{2}}{b_{{0}}}^{
2}-21\, \left( \ln  \left( A \right)  \right) ^{2}{b}^{2}csa_{{3}}{b_{
{0}}}^{2}-9\, \left( \ln  \left( A \right)  \right) ^{2}{c}^{3}qa_{{2}
}b_{{0}}b_{{1}}+11\, \left( \ln  \left( A \right)  \right) ^{2}{c}^{3}
qa_{{3}}b_{{0}}b_{{1}}\right.\\
&\left. + \left( \ln  \left( A \right)  \right) ^{2}{b}^
{2}csa_{{2}}{b_{{1}}}^{2}+6\, \left( \ln  \left( A \right)  \right) ^{
2}{a}^{2}bra_{{2}}{b_{{0}}}^{2}-21\, \left( \ln  \left( A \right) 
 \right) ^{2}{a}^{2}bra_{{3}}{b_{{0}}}^{2}+ \left( \ln  \left( A
 \right)  \right) ^{2}{a}^{2}bra_{{2}}{b_{{1}}}^{2}+1/2\,ap{a_{{2}}}^{
2}b_{{0}}\right. \\
&\left. -2\,\nu\,a_{{3}}b_{{0}}b_{{1}}+ \left( \ln  \left( A \right) 
 \right) ^{2}{c}^{3}qa_{{2}}{b_{{1}}}^{2}+apa_{{0}}a_{{3}}b_{{1}}+apa_
{{1}}a_{{2}}b_{{1}}+apa_{{1}}a_{{3}}b_{{0}}\right. \\
&\left. +6\, \left( \ln  \left( A
 \right)  \right) ^{2}{c}^{3}qa_{{2}}{b_{{0}}}^{2}-21\, \left( \ln 
 \left( A \right)  \right) ^{2}{c}^{3}qa_{{3}}{b_{{0}}}^{2}-9\,
 \left( \ln  \left( A \right)  \right) ^{2}{a}^{2}bra_{{2}}b_{{0}}b_{{
1}}+11\, \left( \ln  \left( A \right)  \right) ^{2}{a}^{2}bra_{{3}}b_{
{0}}b_{{1}}\right. \\
&\left. -9\, \left( \ln  \left( A \right)  \right) ^{2}{b}^{2}csa_{
{2}}b_{{0}}b_{{1}}+11\, \left( \ln  \left( A \right)  \right) ^{2}{b}^
{2}csa_{{3}}b_{{0}}b_{{1}} \right)  P  ^{4}\left( \omega 
 \right)\\
&+ \left( apa_{{0}}a_{{2}}b_{{1}}+apa_{{0}}a_{{3}}b_{{0}}+
apa_{{1}}a_{{2}}b_{{0}}+9\, \left( \ln  \left( A \right)  \right) ^{2}
{c}^{3}qa_{{3}}{b_{{0}}}^{2}-10\, \left( \ln  \left( A \right) 
 \right) ^{2}{c}^{3}qa_{{2}}{b_{{0}}}^{2}- \left( \ln  \left( A
 \right)  \right) ^{2}{c}^{3}qa_{{0}}{b_{{1}}}^{2}+2\, \left( \ln 
 \left( A \right)  \right) ^{2}{c}^{3}qa_{{1}}{b_{{0}}}^{2}\right. \\
&\left. +3\, \left( \ln  \left( A \right)  \right) ^{2}{a}^{2}bra_{{2}}b_{{0}}b_{{
1}}+3\, \left( \ln  \left( A \right)  \right) ^{2}{b}^{2}csa_{{2}}b_{{0
}}b_{{1}}+ \left( \ln  \left( A \right)  \right) ^{2}{a}^{2}bra_{{1}}b
_{{0}}b_{{1}}+ \left( \ln  \left( A \right)  \right) ^{2}{b}^{2}csa_{{
1}}b_{{0}}b_{{1}}\right. \\
&\left. -2\, \left( \ln  \left( A \right)  \right) ^{2}{a}^{2
}bra_{{0}}b_{{0}}b_{{1}}-2\, \left( \ln  \left( A \right)  \right) ^{2
}{b}^{2}csa_{{0}}b_{{0}}b_{{1}}+1/2\,ap{a_{{1}}}^{2}b_{{1}}-2\,\nu\,a_
{{2}}b_{{0}}b_{{1}}-K{b_{{1}}}^{3}+9\, \left( \ln  \left( A \right) 
 \right) ^{2}{a}^{2}bra_{{3}}{b_{{0}}}^{2}\right. \\
&\left. +9\, \left( \ln  \left( A
 \right)  \right) ^{2}{b}^{2}csa_{{3}}{b_{{0}}}^{2}+3\, \left( \ln 
 \left( A \right)  \right) ^{2}{c}^{3}qa_{{2}}b_{{0}}b_{{1}}-10\,
 \left( \ln  \left( A \right)  \right) ^{2}{b}^{2}csa_{{2}}{b_{{0}}}^{
2}-10\, \left( \ln  \left( A \right)  \right) ^{2}{a}^{2}bra_{{2}}{b_{
{0}}}^{2}+2\, \left( \ln  \left( A \right)  \right) ^{2}{a}^{2}bra_{{1
}}{b_{{0}}}^{2}\right. \\
&\left. +2\, \left( \ln  \left( A \right)  \right) ^{2}{b}^{2}c
sa_{{1}}{b_{{0}}}^{2}- \left( \ln  \left( A \right)  \right) ^{2}{a}^{
2}bra_{{0}}{b_{{1}}}^{2}- \left( \ln  \left( A \right)  \right) ^{2}{b
}^{2}csa_{{0}}{b_{{1}}}^{2}+ \left( \ln  \left( A \right)  \right) ^{2
}{c}^{3}qa_{{1}}b_{{0}}b_{{1}}\right.\\
&\left. -2\, \left( \ln  \left( A \right) 
 \right) ^{2}{c}^{3}qa_{{0}}b_{{0}}b_{{1}}-\nu\,a_{{1}}{b_{{1}}}^{2}-
\nu\,a_{{3}}{b_{{0}}}^{2} \right)   P ^{3} \left( \omega  \right)\\
&+ \left( -\nu\,a_{{0}}{b_{{1}}}^{2}-\nu\,a_{{2}}{b_{{0}}}
^{2}-3\, \left( \ln  \left( A \right)  \right) ^{2}{a}^{2}bra_{{1}}{b_
{{0}}}^{2}+4\, \left( \ln  \left( A \right)  \right) ^{2}{a}^{2}bra_{{
2}}{b_{{0}}}^{2}-3\, \left( \ln  \left( A \right)  \right) ^{2}{b}^{2}
csa_{{1}}{b_{{0}}}^{2}+ \left( \ln  \left( A \right)  \right) ^{2}{a}^
{2}bra_{{0}}{b_{{1}}}^{2}\right. \\
&\left.+ \left( \ln  \left( A \right)  \right) ^{2}{
b}^{2}csa_{{0}}{b_{{1}}}^{2} +4\, \left( \ln  \left( A \right) 
 \right) ^{2}{b}^{2}csa_{{2}}{b_{{0}}}^{2}+3\, \left( \ln  \left( A
 \right)  \right) ^{2}{c}^{3}qa_{{0}}b_{{0}}b_{{1}}- \left( \ln 
 \left( A \right)  \right) ^{2}{c}^{3}qa_{{1}}b_{{0}}b_{{1}}-3\,Kb_{{0
}}{b_{{1}}}^{2}+1/2\,ap{a_{{1}}}^{2}b_{{0}}\right. \\
&\left. -2\,\nu\,a_{{1}}b_{{0}}b_{{
1}}+ \left( \ln  \left( A \right)  \right) ^{2}{c}^{3}qa_{{0}}{b_{{1}}
}^{2}+apa_{{0}}a_{{1}}b_{{1}}+apa_{{0}}a_{{2}}b_{{0}}-3\, \left( \ln 
 \left( A \right)  \right) ^{2}{c}^{3}qa_{{1}}{b_{{0}}}^{2}+4\,
 \left( \ln  \left( A \right)  \right) ^{2}{c}^{3}qa_{{2}}{b_{{0}}}^{2
}+3\, \left( \ln  \left( A \right)  \right) ^{2}{a}^{2}bra_{{0}}b_{{0}
}b_{{1}}\right. \\
&\left.- \left( \ln  \left( A \right)  \right) ^{2}{a}^{2}bra_{{1}}b_
{{0}}b_{{1}}
 +3\, \left( \ln  \left( A \right)  \right) ^{2}{b}^{2}csa_
{{0}}b_{{0}}b_{{1}}- \left( \ln  \left( A \right)  \right) ^{2}{b}^{2}
csa_{{1}}b_{{0}}b_{{1}} \right) P^{2} \left( \omega  \right) 
\\
&+ \left( -\nu\,a_{{1}}{b_{{0}}}^{2}- \left( \ln  \left( A \right) 
 \right) ^{2}{c}^{3}qa_{{0}}b_{{0}}b_{{1}}+ \left( \ln  \left( A
 \right)  \right) ^{2}{a}^{2}bra_{{1}}{b_{{0}}}^{2}+ \left( \ln 
 \left( A \right)  \right) ^{2}{b}^{2}csa_{{1}}{b_{{0}}}^{2}-3\,K{b_{{0
}}}^{2}b_{{1}}+1/2\,ap{a_{{0}}}^{2}b_{{1}}\right. \\
&\left. -2\,\nu\,a_{{0}}b_{{0}}b_{{1
}}+ \left( \ln  \left( A \right)  \right) ^{2}{c}^{3}qa_{{1}}{b_{{0}}}
^{2}+apa_{{0}}a_{{1}}b_{{0}}- \left( \ln  \left( A \right)  \right) ^{
2}{a}^{2}bra_{{0}}b_{{0}}b_{{1}}- \left( \ln  \left( A \right) 
 \right) ^{2}{b}^{2}csa_{{0}}b_{{0}}b_{{1}} \right) P \left( \omega
 \right) \\ 
&+1/2\,ap{a_{{0}}}^{2}b_{{0}}-K{b_{{0}}}^{3}-\nu\,a_{{0}}{b_{{0
}}}^{2}=0 \label{kpp}
	\end{aligned}
\end{equation}}
Solving algebraic equations derived by equation the coefficients of the powers of $P(\omega)$ in the previous equation for $a_0$, $a_1$, $a_2$, $a_3$, $b_0$, $b_1$ and $\nu $ gives
{\scriptsize
\begin{equation}
	\begin{aligned}
	 a_0&=0  \\
	 a_1&=\frac{1}{ap} \times \left( b_{{1}} \left( -{a}^{2}br \left( \ln  \left( A \right) 
 \right) ^{2}-{b}^{2}cs \left( \ln  \left( A \right)  \right) ^{2}-{c}
^{3}q \left( \ln  \left( A \right)  \right) ^{2}\right. \right. \\
&\left. \left. \pm \sqrt {{a}^{4}{b}^{2}
{r}^{2} \left( \ln  \left( A \right)  \right) ^{4}+2\,{a}^{2}{b}^{3}cr
s \left( \ln  \left( A \right)  \right) ^{4}+2\,{a}^{2}b{c}^{3}qr
 \left( \ln  \left( A \right)  \right) ^{4}+{b}^{4}{c}^{2}{s}^{2}
 \left( \ln  \left( A \right)  \right) ^{4}+2\,{b}^{2}{c}^{4}qs
 \left( \ln  \left( A \right)  \right) ^{4}+{c}^{6}{q}^{2} \left( \ln 
 \left( A \right)  \right) ^{4}-2\,Kap} \right)  \right)
 \\
	 a_2&= 12\,{\frac { \left( \ln  \left( A \right)  \right) ^{2}b_{{1}} \left( 
{a}^{2}br+{b}^{2}cs+{c}^{3}q \right) }{ap}}
 \\
a_3&=-12\,{\frac { \left( \ln  \left( A \right)  \right) ^{2}b_{{1}} \left( 
{a}^{2}br+{b}^{2}cs+{c}^{3}q \right) }{ap}}
\\
	 b_0&=0\\
	 %b_1&= \\
	\nu&=\pm \sqrt {{a}^{4}{b}^{2}{r}^{2} \left( \ln  \left( A \right)  \right) ^{
4}+2\,{a}^{2}{b}^{3}crs \left( \ln  \left( A \right)  \right) ^{4}+2\,
{a}^{2}b{c}^{3}qr \left( \ln  \left( A \right)  \right) ^{4}+{b}^{4}{c
}^{2}{s}^{2} \left( \ln  \left( A \right)  \right) ^{4}+2\,{b}^{2}{c}^
{4}qs \left( \ln  \left( A \right)  \right) ^{4}+{c}^{6}{q}^{2}
 \left( \ln  \left( A \right)  \right) ^{4}-2\,Kap}
\\
	\end{aligned}
\end{equation}}
where $ap \neq 0$ for arbitrarily chosen $a$, $b$, $c$, $K$ and $b_1$. Thus, the solutions to (\ref{zkode}) are expressed as
{\tiny
\begin{equation}
\begin{aligned}
	U_{3,4}(\omega)&=\frac{1}{b_1\left(\frac{1}{1+\tilde{d}A^{\omega}}\right)} \times \left[\frac{1}{ap} \times \left( b_{{1}} \left( -{a}^{2}br \left( \ln  \left( A \right) 
 \right) ^{2}-{b}^{2}cs \left( \ln  \left( A \right)  \right) ^{2}-{c}
^{3}q \left( \ln  \left( A \right)  \right) ^{2}+\nu \right)  \right)\times \left(\frac{1}{1+\tilde{d}A^{\omega}}\right) \right. \\
&\left. +12\,{\frac { \left( \ln  \left( A \right)  \right) ^{2}b_{{1}} \left( 
{a}^{2}br+{b}^{2}cs+{c}^{3}q \right) }{ap}}\times \left(\frac{1}{1+\tilde{d}A^{\omega}}\right)^2 \right.\\
&\left.-12\,{\frac { \left( \ln  \left( A \right)  \right) ^{2}b_{{1}} \left( 
{a}^{2}br+{b}^{2}cs+{c}^{3}q \right) }{ap}}\times \left(\frac{1}{1+\tilde{d}A^{\omega}}\right)^3\right] \\
&= \left[\frac{1}{ap} \times \left(-{a}^{2}br \left( \ln  \left( A \right) 
 \right) ^{2}-{b}^{2}cs \left( \ln  \left( A \right)  \right) ^{2}-{c}
^{3}q \left( \ln  \left( A \right)  \right) ^{2}+\nu  \right) \right. \\
&\left. +12\,{\frac { \left( \ln  \left( A \right)  \right) ^{2}\left( 
{a}^{2}br+{b}^{2}cs+{c}^{3}q \right) }{ap}}\times \left(\frac{1}{1+\tilde{d}A^{\omega}}\right) \right.\\
&\left.-12\,{\frac { \left( \ln  \left( A \right)  \right) ^{2} \left( 
{a}^{2}br+{b}^{2}cs+{c}^{3}q \right) }{ap}}\times \left(\frac{1}{1+\tilde{d}A^{\omega}}\right)^2\right] \\
\end{aligned}
\end{equation}}
for arbitrarily chosen $b_1$, $K$, $a$, $b$ and $c$. The return to the original variables gives the solutions as
{\tiny
\begin{equation}
\begin{aligned}
	u_{3,4}(x,y,z,t)&=\left[\frac{1}{ap} \times   \left( -{a}^{2}br \left( \ln  \left( A \right) 
 \right) ^{2}-{b}^{2}cs \left( \ln  \left( A \right)  \right) ^{2}-{c}
^{3}q \left( \ln  \left( A \right)  \right) ^{2}+ \nu  \right) \right. \\
&\left. +12\,{\frac { \left( \ln  \left( A \right)  \right) ^{2} \left( 
{a}^{2}br+{b}^{2}cs+{c}^{3}q \right) }{ap}}\times \left(\frac{1}{1+\tilde{d}A^{ax+by+cz-\nu \frac{t^{\beta}}{\beta}}}\right) \right.\\
&\left.-12\,{\frac { \left( \ln  \left( A \right)  \right) ^{2}b_{{1}} \left( 
{a}^{2}br+{b}^{2}cs+{c}^{3}q \right) }{ap}}\times \left(\frac{1}{1+\tilde{d}A^{ax+by+cz-\nu \frac{t^{\beta}}{\beta}}}\right)^2\right] \\
\end{aligned}
\end{equation}}
where 
{\scriptsize
\begin{equation}
	\nu=\pm \sqrt {{a}^{4}{b}^{2}{r}^{2} \left( \ln  \left( A \right)  \right) ^{
4}+2\,{a}^{2}{b}^{3}crs \left( \ln  \left( A \right)  \right) ^{4}+2\,
{a}^{2}b{c}^{3}qr \left( \ln  \left( A \right)  \right) ^{4}+{b}^{4}{c
}^{2}{s}^{2} \left( \ln  \left( A \right)  \right) ^{4}+2\,{b}^{2}{c}^
{4}qs \left( \ln  \left( A \right)  \right) ^{4}+{c}^{6}{q}^{2}
 \left( \ln  \left( A \right)  \right) ^{4}-2\,Kap}
\end{equation}}
The algebraic system (\ref{kpp}) has some more solutions written as
{\tiny
\begin{equation}
	\begin{aligned}
	 a_0&={\frac { \left( - \left( \ln  \left( A \right)  \right) ^{2}{a}^{2}brb
_{{1}}- \left( \ln  \left( A \right)  \right) ^{2}{b}^{2}csb_{{1}}-
 \left( \ln  \left( A \right)  \right) ^{2}{c}^{3}qb_{{1}}+b_1\nu
 \right) b_{{0}}}{apb_{{1}}}}
  \\
	 a_1&={\frac {12\, \left( \ln  \left( A \right)  \right) ^{2}{a}^{2}brb_{{0}
}- \left( \ln  \left( A \right)  \right) ^{2}{a}^{2}brb_{{1}}+12\,
 \left( \ln  \left( A \right)  \right) ^{2}{b}^{2}csb_{{0}}- \left( 
\ln  \left( A \right)  \right) ^{2}{b}^{2}csb_{{1}}+12\, \left( \ln 
 \left( A \right)  \right) ^{2}{c}^{3}qb_{{0}}- \left( \ln  \left( A
 \right)  \right) ^{2}{c}^{3}qb_{{1}}+b_1 \nu}{ap}} \\
	 a_2&= -12\,{\frac { \left( {a}^{2}brb_{{0}}-{a}^{2}brb_{{1}}+{b}^{2}csb_{{0}
}-{b}^{2}csb_{{1}}+{c}^{3}qb_{{0}}-{c}^{3}qb_{{1}} \right)  \left( 
\ln  \left( A \right)  \right) ^{2}}{ap}}
 \\
a_3&=-12\,{\frac { \left( \ln  \left( A \right)  \right) ^{2}b_{{1}}
 \left( {a}^{2}br+{b}^{2}cs+{c}^{3}q \right) }{ap}}
\\
	 %b_0&=0\\
	 %b_1&= \\
	\nu&=\pm { {\sqrt { \left( \ln  \left( A \right)  \right) ^{4}{a}^{4}{b}^{
2}{r}^{2}+2\, \left( \ln  \left( A \right)  \right) ^{4}{
a}^{2}{b}^{3}crs+2\, \left( \ln  \left( A \right) 
 \right) ^{4}{a}^{2}b{c}^{3}qr+ \left( \ln  \left( A
 \right)  \right) ^{4}{b}^{4}{c}^{2}{s}^{2}+2\, \left( 
\ln  \left( A \right)  \right) ^{4}{b}^{2}{c}^{4}qs+
 \left( \ln  \left( A \right)  \right) ^{4}{c}^{6}{q}^{2}
-2\,Kap}}}
	\end{aligned}
\end{equation}}
where $b_0$, $b_1$, $K$, $a$, $b$ and $c$ are arbitrary constants. Thus, the solutions to (\ref{zkode}) are obtained as

{\tiny
\begin{equation}
\begin{aligned}
	U_{5,6}(\omega)&=\left( \frac{1}{b_0+b_1\times \left(\frac{1}{1+\tilde{d}A^{\omega}}\right)}\right) \times \left[ {\frac { \left( - \left( \ln  \left( A \right)  \right) ^{2}{a}^{2}br- \left( \ln  \left( A \right)  \right) ^{2}{b}^{2}cs-
 \left( \ln  \left( A \right)  \right) ^{2}{c}^{3}q+\nu
 \right) b_{{0}}}{ap}}\right. \\
	&\left. +\left({\frac {12\, \left( \ln  \left( A \right)  \right) ^{2}{a}^{2}brb_{{0}
}- \left( \ln  \left( A \right)  \right) ^{2}{a}^{2}brb_{{1}}+12\,
 \left( \ln  \left( A \right)  \right) ^{2}{b}^{2}csb_{{0}}- \left( 
\ln  \left( A \right)  \right) ^{2}{b}^{2}csb_{{1}}+12\, \left( \ln 
 \left( A \right)  \right) ^{2}{c}^{3}qb_{{0}}- \left( \ln  \left( A
 \right)  \right) ^{2}{c}^{3}qb_{{1}}+b_1 \nu}{ap}}\right) \times \left(\frac{1}{1+\tilde{d}A^{\omega}}\right)\right. \\
	&\left. +\left( -12\,{\frac { \left( {a}^{2}brb_{{0}}-{a}^{2}brb_{{1}}+{b}^{2}csb_{{0}
}-{b}^{2}csb_{{1}}+{c}^{3}qb_{{0}}-{c}^{3}qb_{{1}} \right)  \left( 
\ln  \left( A \right)  \right) ^{2}}{ap}}\right)\times \left(\frac{1}{1+\tilde{d}A^{\omega}}\right)^2 \right. \\ 
	&\left. +\left( -12\,{\frac { \left( \ln  \left( A \right)  \right) ^{2}b_{{1}}
 \left( {a}^{2}br+{b}^{2}cs+{c}^{3}q \right) }{ap}}\right)\times\left(\frac{1}{1+\tilde{d}A^{\omega}}\right)^3 \right]
\end{aligned}
\end{equation}}
for arbitrarily chosen $b_0$, $b_1$, $K$, $a$, $b$ and $c$ and
{\tiny 
\begin{equation}
		\nu=\pm { {\sqrt { \left( \ln  \left( A \right)  \right) ^{4}{a}^{4}{b}^{
2}{r}^{2}+2\, \left( \ln  \left( A \right)  \right) ^{4}{
a}^{2}{b}^{3}crs+2\, \left( \ln  \left( A \right) 
 \right) ^{4}{a}^{2}b{c}^{3}qr+ \left( \ln  \left( A
 \right)  \right) ^{4}{b}^{4}{c}^{2}{s}^{2}+2\, \left( 
\ln  \left( A \right)  \right) ^{4}{b}^{2}{c}^{4}qs+
 \left( \ln  \left( A \right)  \right) ^{4}{c}^{6}{q}^{2}
-2\,Kap}}}
\end{equation}}
Returning the original variables gives the solution to the comformable time fractional ZK equation (\ref{cfzk}) as
{\tiny
\begin{equation}
\begin{aligned}
	&u_{5,6}(x,y,z,t)=\left( \frac{1}{b_0+b_1\times \left(\frac{1}{1+\tilde{d}A^{ax+by+cz-\nu\frac{t^{\beta}}{\beta}}}\right)}\right) \times \left[ {\frac { \left( - \left( \ln  \left( A \right)  \right) ^{2}{a}^{2}br- \left( \ln  \left( A \right)  \right) ^{2}{b}^{2}cs-
 \left( \ln  \left( A \right)  \right) ^{2}{c}^{3}q+\nu
 \right) b_{{0}}}{ap}}\right. \\
	&\left. +\left({\frac {12\, \left( \ln  \left( A \right)  \right) ^{2}{a}^{2}brb_{{0}
}- \left( \ln  \left( A \right)  \right) ^{2}{a}^{2}brb_{{1}}+12\,
 \left( \ln  \left( A \right)  \right) ^{2}{b}^{2}csb_{{0}}- \left( 
\ln  \left( A \right)  \right) ^{2}{b}^{2}csb_{{1}}+12\, \left( \ln 
 \left( A \right)  \right) ^{2}{c}^{3}qb_{{0}}- \left( \ln  \left( A
 \right)  \right) ^{2}{c}^{3}qb_{{1}}+b_1 \nu}{ap}}\right) \times \left(\frac{1}{1+\tilde{d}A^{ax+by+cz-\nu\frac{t^{\beta}}{\beta}}}\right)\right. \\
	&\left. +\left( -12\,{\frac { \left( {a}^{2}brb_{{0}}-{a}^{2}brb_{{1}}+{b}^{2}csb_{{0}
}-{b}^{2}csb_{{1}}+{c}^{3}qb_{{0}}-{c}^{3}qb_{{1}} \right)  \left( 
\ln  \left( A \right)  \right) ^{2}}{ap}}\right)\times \left(\frac{1}{1+\tilde{d}A^{ax+by+cz-\nu\frac{t^{\beta}}{\beta}}}\right)^2 \right. \\ 
	&\left. +\left( -12\,{\frac { \left( \ln  \left( A \right)  \right) ^{2}b_{{1}}
 \left( {a}^{2}br+{b}^{2}cs+{c}^{3}q \right) }{ap}}\right)\times\left(\frac{1}{1+\tilde{d}A^{ax+by+cz-\nu\frac{t^{\beta}}{\beta}}}\right)^3 \right]
\end{aligned}
\end{equation}}
where
{\tiny 
\begin{equation}
		\nu=\pm { {\sqrt { \left( \ln  \left( A \right)  \right) ^{4}{a}^{4}{b}^{
2}{r}^{2}+2\, \left( \ln  \left( A \right)  \right) ^{4}{
a}^{2}{b}^{3}crs+2\, \left( \ln  \left( A \right) 
 \right) ^{4}{a}^{2}b{c}^{3}qr+ \left( \ln  \left( A
 \right)  \right) ^{4}{b}^{4}{c}^{2}{s}^{2}+2\, \left( 
\ln  \left( A \right)  \right) ^{4}{b}^{2}{c}^{4}qs+
 \left( \ln  \left( A \right)  \right) ^{4}{c}^{6}{q}^{2}
-2\,Kap}}}
\end{equation}}

\section{Conclusion}
In the study, the generalized Kudryashov method is implemented to some conformable time fractional PDEs defined in three space dimensions, name the conformable time fractional JM and ZK equations. The compatible wave transform has a significant role in the solutions steps. Reducing both equations to some ODEs and implementation of the generalized form of the Kudryashov method derive some explicit exact solutions to them. These explicit solutions can be represented in rational forms of some finite exponential function series.

{\textbf{Acknowledgement:} A part of this study was presented orally in International  Congress on Fundamental and Applied Sciences 2017, Sarajevo, Bosnia and Herzegovina.}

\end{document}